# Nano-size fragmentation of Tantalum in Copper composite using additive manufacturing

*Rakesh Das[1,§], Pawan Kumar Dubey[2,§], Raphael Benjamim de Oliveira[3], Douglas S. Galvao[4], Indranil Manna[1], Sameehan S. Joshi[5,6], Peter Samora Owuor[7], Leonardo D. Machado[3,\*], Nirmal Kumar Katiyar[8,\*], Suman Chakraborty[2,\*], Chandra Sekhar Tiwary[1,\*]*

[1]*Department of Metallurgical and Materials Engineering, Indian Institute of Technology Kharagpur, Kharagpur-721302, West Bengal, India*

[2]*Department of Mechanical Engineering, Indian Institute of Technology Kharagpur, Kharagpur-721302, West Bengal, India*

[3]*Departamento de Física Teórica e Experimental, Universidade Federal do Rio Grande do Norte, Natal, Rio Grande do Norte 59072-970, BRAZIL*

[4]*State University of Campinas, Applied Physics Department and Center for Computing Engineering and Sciences, Campinas, São Paulo, 13083-970, Brazil*

[5]*Materials Science and Engineering, University of North Texas, Denton Texas, USA, 76207.*

[6]*Center for Agile and Adaptive Additive Manufacturing, University of North Texas, Denton Texas, USA, 76207.*

[7]*Carbon Center of Excellence Morgan Advanced Materials, State College, Pennsylvania, 16803, USA*

[8]*Amity Institute of Applied Sciences, Amity University Noida 201303, India*

*\*Corresponding authors E-mail:*

*Leonardo D. Machado: leonardo@fisica.ufrn.br,*

*Suman Chakraborty: suman@mech.iitkgp.ac.in,*

*Nirmal Kumar Katiyar: nirkatiyar@gmail.com,*

*Chandra Sekhar Tiwary: chandra.tiwary@metal.iitkgp.ac.in,*

*§ Equal contribution*

**Abstract**

The biggest challenge in manufacturing an immiscible system is phase segregation and non-uniformity inside the composite matrix. Additive manufacturing has the potential to overcome these difficulties due to the high cooling rate achieved during the process. Here we have developed immiscible Copper-based composites reinforced with Tantalum, which were fabricated using the powder bed fusion melting (PBF-M) technique. The distinct advantage of utilizing Tantalum in this process resides in its high melting point, allowing it to remain in particle form within the composite and contribute to its mechanical and surface/wear properties. The PBF-M results in the in situ fragmentation of micron-size Tantalum particles into nanoparticle form through a surface roughening process during laser interaction, enhancing its mechanical and wear properties. The microstructural evolution of Cu-Ta composites is explained through multiscale numerical modeling. The enhanced yield strength and the dynamics of the Ta particles were corroborated by molecular dynamics simulations. The maximum yield strength is exhibited by Cu-5wt%Ta of 80 MPa. Addition of Ta also have significant improvement in wear properties of composites. The current results can be exploited to develop complex shape, high energy efficient copper-based composites.

**Keyword**s: Additive manufacturing, microstructure, mechanical properties, molecular Dynamics, computational fluid dynamics.

# 1. Introduction

Fusion-based additive manufacturing, e.g., powder bed fusion (PBF) and direct energy deposition (DED), has revolutionized the fields of materials processing, enabling the fabrication of intricate 3D objects with micrometer resolution through the precise layer-by-layer melting/solidification[1–3]. These advancements have found significant applications in industries such as automobiles and aerospace[4]. The processing of pure metals is simple and well-controlled; however, they show inferior mechanical properties. Therefore, mixing with other elements is essential to improving pure metals' structural and functional properties [5–7]. The addition of reinforcement in metals with controlled composition and microstructure remains one of the key challenges in the current Fusion-based additive manufacturing process. Hence, several efforts have been made to tune the microstructure and mechanical properties using additive manufacturing (AM) in the last few years. However, significant challenges arise when dealing with immiscible elements with very different melting temperatures and densities. The main challenge in such a composite system is phase/particle segregation, which results in poor mechanical properties. The phase separation of the immiscible elements increases during conventional melting/solidification processes involving a slow rate of cooling.

Powder bed fusion melting (PBF-M) processes involve a convergent source of heat (focus laser), which provides a rapid solidification (heating and cooling) process[8]. Such a high cooling rate does not allow the immiscible elements to diffuse through the liquid metal, which results in uniform distribution of the second phase/particle in the metal matrix and also prevents the oxidation of constituent elements in the inert atmosphere. The PBF-M processes can be more advantageous to immiscible elements with different melting temperatures, large density differences, and high heat of formation values[9], such as FCC Copper (Cu) and BCC metals, e.g., Tantalum (Ta). As Cu is a very good thermal conductor, it melts quickly under a focus laser beam and dissipates heat rapidly, which does not allow phase segregation. The lower absorptivity of Cu powders through high wavelength laser can be increased through the addition of secondary reinforcement with the optimization of laser parameters. Also, rapid solidification conditions can be achieved through optimization such that the immiscible (high-temperature BCC metal) particle/phase can retain its original position in the metal matrix.

The utilization of AM for Cu-based composites has found significant applications in electrical and thermal fields[10,11]. Therefore, the fabrication process necessitates to enhance specific strength, as well as thermal and electrical conductivity. Consequently, extensive efforts have been devoted to developing Cu-based composites with improved high strength and wear resistance while preserving their electrical and thermal properties[12–15]. W-Cu composite, fabricated through the PBF technique, is the most well-known immiscible system that has been extensively reported[16]. Vüllers et al. conducted a study on Cu-W thin films using sputter deposition. By carefully controlling the deposition process, they enhanced the wear resistance of the Cu-W thin films without compromising their thermal conductivity[17]. Darling et al. investigated the immiscible system of Cu and Ta, utilizing high-energy ball milling to prepare the system, followed by consolidation through high-temperature equal-channel angular extrusion (ECAE)[18]. The initial nanocrystalline Cu material was susceptible to coarsening at low temperatures, leading to the loss of its nanostructured properties during processing. However, adding Ta particles (with a size of approximately 70 nm) can effectively hinder the coarsening behavior and prevent the loss of the nanocrystalline structure during coarsening and other related processes[19]. Similarly, Venugopal et al. reported that the Cu-Ta nanocomposite compressive strength was 3.6-5 times higher than micro composites[20]. Frolov et al. conducted a study on Cu-6.5Ta and demonstrated that the segregation of Ta along the grain boundaries reduces grain growth[21]. In a recent study, immiscible Cu-hexagonal Boron Nitride and Cu-Graphene composites were fabricated through the SLM process[22]. Their study demonstrated the maximum graphene content that can be added to fabricate Cu-Gr composite utilizing the PBF-M method successfully.

Out of various strengthening mechanisms, dispersion strengthening and grain boundary strengthening can be achieved through the PBF-M process [23]. However, it is worth mentioning that while finer grains contribute to strengthening, they can also reduce electrical and thermal conductivity [24]. The grain boundaries in these composites are characterized by a high density of defects, such as vacancies and dangling bonds, which can significantly influence electronic and thermal transport properties. Therefore, the engineering of interfaces is crucial for the smooth transport of electrons across grain boundaries is a crucial step. Recently, several theoretical studies have been reported on immiscible systems, mostly involving MD simulations that focus on phase evolution and mechanical properties of Cu and Ta interface at different lengths [25–28]. The simulations reveal that although Ta is thermodynamically immiscible with Cu, it shows a strong tendency to immerse into Cu at the atomic scale [27]. Also, it has been found that amorphous structures mainly form on the Cu layers, and a small amount of the BCC Ta particles transform into amorphous[28]. Thus, this misfit energy could act as a source of dislocation and barriers to preclude the motion of the dislocations[29]. Previous MD simulations on Cu-Ta composites have investigated their structural stability, mechanical strength, and the diffusion of atoms at the interface[30–32].

In this study, we have fabricated Cu-Ta composites through an optimized PBF-M method and investigated the reliability and feasibility of PBF-M in manufacturing immiscible systems. As the LPBF process provides a higher cooling rate, the entrapment of dispersoids at the same location in the composite can be achieved. Additionally, the study aims to analyze the influence of Ta on the strength of the materials fabricated using PBF-M additive manufacturing. The experimental observations of PBF-M, i.e., mechanical strength due to Ta and decrease in Ta particle size inside the Cu-matrix and their interface, were further investigated and validated using MD simulations. Also, the mechanical behavior and strain hardening, along with dislocation interaction, have been explained based on the MD simulations.

## 2. Experimental Details

### 2.1. Materials and sample preparation

The Cu and Ta powders utilized in this study were obtained from Thermo Fisher® and possessed a purity of 99.9%. The size of the powder particles was determined through scanning electron microscopy (SEM) (Jeol JSM-IT300HR). Five distinct samples were prepared, each containing varying amounts of Ta denoted as Cu-X wt% Ta, where X represents 0, 0.5, 1, 2, and 5. The composition of Cu without Ta is designated as PBF-M Cu. To ensure a homogeneous mixture of powders, the Cu and Ta powders were blended in a Tumbler machine for 45 minutes at a rotational speed of 60 rpm. After blending the compositions, each mixture was subjected to compaction using a hydraulic press to prepare the pellet. A cylindrical die and punch system with a diameter of 25 mm was used for the pellet fabrication, and a load of 5 tons was applied for each pellet fabrication. This load was applied during the compaction to only hold all powders in place so that powder dispersion would not happen during laser melting during the laser PBF-M process. The final height of the compacted samples ranged from 4 mm to 4.5 mm. The subsequent experiment involved the PBF-M technique, which was carried out using a continuous Fiber Diode laser machine (LDF 6000-40) with a laser spot diameter of 3.6 mm. A power output of 800 W and the scanning speed of 40 mm/s were maintained for all the Cu-Ta composites. The laser was scanned over the Cu-Ta pellets' surface in a raster pattern, and the top view of laser path movement is schematically shown in Figure 1(a). The Cu-Ta pellets were melted in a single pass by the focus laser. The optimum processing parameter (laser power and scan rate) is based upon the repeated trial experiments performed on the Cu pellet alone to know the laser depth and the Cu melting behavior. In general, the composites have to endure delamination and distortion during the PBF-M process. In order to minimize the delamination and distortion, we have optimized the laser energy and scan speed on pure Cu. We have used the optimized laser processing parameters to study the Cu-Ta composite behavior with different Ta concentrations, and to keep the results brief, we have provided only the results of the optimized processing conditions. An SS 316L steel substrate was employed to enable convenient product removal, and the pellets were placed over 316L without any assistance. The entire process took place in an argon atmosphere with an oxygen concentration below 100 ppm to minimize the influence of oxygen on the final product.

## 2.2. Characterizations

Each sample underwent polishing and etching procedures following the laser melting to examine its microstructure. Samples were polished along longitudinal (normal to laser beam direction) and transverse (along laser beam direction) directions, first with emery papers of different grades, starting from a coarse (grade 240) paper and moving to a finer one (grade 2000) in steps and then a colloidal suspension of diamond was used for final polishing. For the microstructure analysis, both optical and scanning electron microscopes were employed. The polished samples were examined using a Leica-DM2500M optical microscope. Potassium dichromate was utilized as an etchant to improve the visibility of grains and other features. The density of the laser-melted pellets was measured by determining the mass and volume of each pellet, and then density was calculated using $\rho = m/V$, where $\rho$ is density, m is mass, and V is the volume of the sample. X-ray diffraction (XRD) analysis was performed using Bruker D8 Advance with Lynx eye detector at a range of 20° to 100° using Cu-k$\alpha$ ($\lambda = 0.1542$ nm) radiation for the phase identification. Micro-Vickers hardness and compression tests were carried out to assess the mechanical properties. The micro-Vickers hardness measurements were performed using a UHL VMHT MOT hardness tester, with a load of 1000 gf and a dwell time of 15 s. Both as-fabricated and heat-treated samples were evaluated using this method. To perform the compression tests, cylindrical samples with a diameter of 3 mm were obtained from each PBF-M-processed Cu-Ta composite. These samples were cut using an Electrical Discharge Machine (EDM), following the guidelines set by ASTM E10004-07 standards, with an aspect ratio (height to diameter ratio) ranging from 1.5 to 17. Compression test samples were cut from the composite along a transverse direction, i.e., in the direction of the laser beam. After the compression tests, the fracture surfaces of the investigated samples were analyzed using a scanning electron microscope (SEM). Energy-dispersive X-ray spectroscopy (EDS) was performed to conduct elemental analysis of the Cu-Ta composites. Tribological properties were assessed using an Anton Paar Tribometer under a load of 5 N for a duration of 15 minutes. A steel ball with a diameter of 10 mm was utilized for the wear test, and the wear surface morphology was analyzed using SEM. The effect of precipitates was investigated using a high-resolution Transmission Electron Microscope (HR-TEM). For TEM analysis, thin plates were cut from each Cu-Ta composite using EDM with a diameter of 3 mm, and each plate was thinned down to 100 μm using silicon carbide (SiC) emery paper. Subsequently, dimpling at the center of these 3 mm discs was performed using a dimpling machine. Then, each sample was cleaned with acetone, and precision ion milling was performed to obtain a transparent electron region for TEM analysis. TEM analysis was conducted using an aberration-corrected Jeol Neoarm 200f microscope operated at 230 kV. Multiple images were acquired in brightfield and high-resolution modes to examine the distribution of Ta particles. The thermal conductivity of the Cu-Ta composites was determined through a high-resolution InfraRed (IR) camera setup. In this setup, the composite samples were placed over a hot plate having a surface temperature of 120 °C. The IR camera was pointed at the sample surface for the temperature measurement at three points. The temperature at three points on the surface of composites was recorded with the increment of time, and qualitatively, the comparison of thermal conductivity was reported based on the difference in temperature at two extreme points.

## 2.3. Numerical Thermofluidic modeling of PBF-M process

The numerical modeling of the PBF-M method incorporates two essential components: the Discrete Element Method (DEM) and Computational Fluid Dynamics (CFD). The model is a one-way coupled CFD-DEM model that uses DEM to create a compressed Cu-Ta pellet using the open-source code LIGHTS. Thereafter, the radii and coordinates of different Cu and Ta particles were imported into ANSYS Fluent to model the laser melting process. A uniform mesh with hexahedral cells of size four μm was chosen for the simulation. The mesh size was chosen so that the spherical shape of the metal particles of the given size (44 μm) could be resolved accurately. DEM captures the microscale powder particles' motion and dynamics, while CFD accounts for the thermo-fluidic transport phenomena. These interconnected models rely on each other, as the arrangement and motion of the particles from DEM influences the CFD results during laser melting and vice versa. The motion of particles due to fluid transport affects the particle motion. Two different configurations for the Cu-Ta pellet were modeled using DEM with 2wt% and 5wt%Ta. Both configurations consist of uniform-sized Cu and Ta particles (20 μm and 50 μm, respectively). The laser parameters used in the simulations are obtained using normalization of "Area Energy Density", which is defined as P/v.d where $P$ is the laser power, $v$ is the scan speed, and $d$ is the laser spot diameter. It represents the energy absorbed by the laser spot per unit of time when moving with velocity v. The method is used for scaling down the dimensions of the pellet. We used this method as the size of the pellet used in the experiment is very large, and modeling such a large system with the required cell size in the mesh would be computationally very expensive. A similar parameter, volumetric energy density, is commonly used in PBF-M works, which employs powder layer thickness and is used to compare two different systems. The laser heat source was applied to the top surface boundary of the pellet. The heat from the laser on the top surface melted the Cu and Ta particles, creating the melt pool. The surface tension gradient and Marangoni convection resulted in the melt pool movement and oscillations. The boundary surfaces of the melt pool were continuously subjected to heat losses due to radiation, convection, and conduction, leading to its rapid solidification. The details of equations, boundary conditions, associated thermophysical properties, and underlying assumptions have been presented in Supplementary Information.

## 2.4. Molecular dynamics simulations

Metallic alloys consisting of immiscible metals have been extensively studied using Molecular dynamics (MD) simulations, and the results are consistent with *ab initio* data and experimental results when the potential EAM ("Embedded Atom Method") is used [33–35]. To gain insights into the experimental results, MD simulations have been performed using EAM potentials [36,37], as implemented in the LAMMPS MD package [38]. The parameterization used in this paper is from X.W. Zhou et al. [37], a widely cited article used in MD simulations of metallic alloys. It was developed to obtain parameterizations for the EAM potential applied in the deposition of multilayers of different types of metals. Its results are in agreement with the data available at the time. Further improvements were made to achieve better reproducibility of experimental results, as reported in Pun, G. P. et al. [39].

The unit cell of our periodic system is composed of a cubic sample of 1024 BCC Ta atoms embedded within 53924 FCC Cu atoms. Thus, Ta atoms represent 5.1% of the total mass. After creating the structural model systems, the energy was minimized by performing 20000 steps of the conjugate gradient algorithm, during which the unit cell size and atomic positions were allowed to vary. After minimization, all three cell parameter vectors had the same length of 8.69 nm. In all simulations, the temperature and pressure were controlled using a chain of three Nosé-Hoover thermostats and barostats (NPT ensemble). In the first type of simulation, we examined how the increase in temperature due to the heating generated by the laser would affect the morphology at a Cu-Ta interface. To this end, the system was set at 300 K, and the temperature was gradually increased to 1000 K over 5.0 x $10^5$ steps. Next, the temperature was kept fixed at 1000 K for 1.0 x $10^6$ steps, and then the temperature was gradually lowered back to 300 K over 5.0 x $10^5$ steps. The temperature profile/distribution during the PBF-M process is complex and can vary depending on several processing conditions, such as powder size, laser energy, scan rate, heat transfer, etc. Simulating actual conditions using MD simulations is challenging due to size effects and other limitations. Hence, in our MD simulation, we analyzed the structures after a thermal equilibration at 1000 K for 1.0 x $10^6$ steps so that the interface of Cu and Ta reaches a minimum energy state followed by mechanical deformation. The temperature selection was based on **Figure 6**, which indicated that 1000 K was a typical temperature during heating, except for regions close to the surface. Although the predicted peak temperatures reached ~1900 K, it remained at that level for ~10 μs. Furthermore, the associated heating and cooling rates were of the order of ~$10^7$ K/s. These thermo-kinetics were insufficient for achieving the effect of a relatively slow thermal treatment such as annealing. In every stage of these simulations, the system was kept relaxed by setting the target pressure of the barostat to zero atm, and a timestep of 0.1 fs was used to stabilize the system. We assumed a Ta atom was at the interface if it was within 2.8 Å of a Cu atom. In the second type of simulation, the system was compressed to analyze the morphology changes due to fracture at a fixed temperature of 300 K. To achieve this goal, the final configuration of the heating simulations was first equilibrated for 5.0 x $10^5$ steps with a target pressure of zero atm. Then, the barostat was turned off along the x direction, and the cell parameter vector was gradually decreased at a rate of 0.001 per picosecond. The system was evaluated for 2.5 x $10^6$ steps of 0.05 fs. MD snapshots were used to evaluate morphology changes near a Ta particle.

3. **Results & Discussions**

3.1. **Microstructure Analysis**

In this study, blending was a very crucial process for the uniform physical mixing of powders and to avoid agglomeration. After blending, we compacted the powders in a die-punch system to avoid powder splashing during laser melting. The objective of the present work was to try to understand the bulk fabrication of an immiscible composite system during the PBF-M processing. For this purpose, we melted the whole pellet through the optimization of laser parameters such as laser power and scan speed. The adopted approach provided a sound way to achieve the desired results. The morphology of Cu and Ta powders in their as-received state is shown in Figure S1(a) and S1(b), respectively. The Cu powder (99% pure) exhibits a dendritic structure with a particle size of ~13.5 μm, whereas the Ta powder (99% pure) consists of a particle size of ~11 μm and the size distribution plot is shown in Figure S1(c) and S1(d). As Cu powders show low absorptivity, a larger diameter of 3.6mm beam diameter was used to melt the pellet. The laser was rastered over the pellet in a predefined path. Five samples were prepared by combining Cu and Ta powders in different ratios, with Cu as the matrix and Ta as the reinforcement. The experimental schematic of the powder mixture, pellet fabrication, and PBF-M process is depicted in Figure S1(e). The density of the processed samples lies in the range of 7.5-7.95 g/cm3. The phase diagram of the Cu-Ta binary system is shown in Figure S2, and within the compositional range, both materials remain unmixed. The laser energy density employed in the process is insufficient to melt the Ta particles, given that the melting temperature of Ta is 3017°C. The melted regions show the absence of defects and voids, as observed in the optical micrographs presented in Figures 1(a) and 1(b). The PBF-M Cu results in equiaxed grain structures throughout the sample. However, as the Ta concentration increases, significant changes in grain morphologies become evident.

The detailed optical and SEM micrographs for longitudinal and transverse directions are shown in **Figure 1**. In the PBF-M process, the evolution of microstructure strongly depends upon temperature gradient (G) and growth rate (R) at the melt pool boundary. Due to the high dissipation of PBF-M Cu, it shows low G/R values, promoting an equiaxed microstructure in the PBF-M Cu. In contrast, the 0.5wt%Ta and 1wt%Ta composite cases show the same features as PBF-M Cu along the longitudinal direction, whereas in the transverse direction, elongated grains can be seen for 0.5wt%Ta as shown in **Figure 1(b), 1(c), 1(g)** and **1(h),** respectively. Cu-2wt%Ta shows the formation of columnar dendritic microstructure in both directions, and Cu-5wt%Ta composites exhibit cellular and larger grain morphology as shown in **Figure 1(d)**, **1(i)**, **1(e), and 1(j)**, respectively. The increasing Ta concentration has a significant effect in altering the G/R values during the PBF-M process. The microstructural analysis along the longitudinal and transverse directions indicates a difference in grain morphology, suggesting anisotropic microstructure in PBF-M processed composites. **Figure S3** depicts the representative SEM micrographs and elemental analysis of 1wt%Ta and 5wt%Ta, respectively. It shows the distribution of Ta inside the Cu matrix after the PBF-M process. The increase in uniform distribution can be visualized when comparing **Figure S3(a)** and **Figure S3(c)**. SEM micrographs reveal the particle size of both composites to be in the range of 300 to 600 nm. The interface between Cu and Ta exhibits no cracks or voids, confirming the sound bonding between the two particles after laser melting.

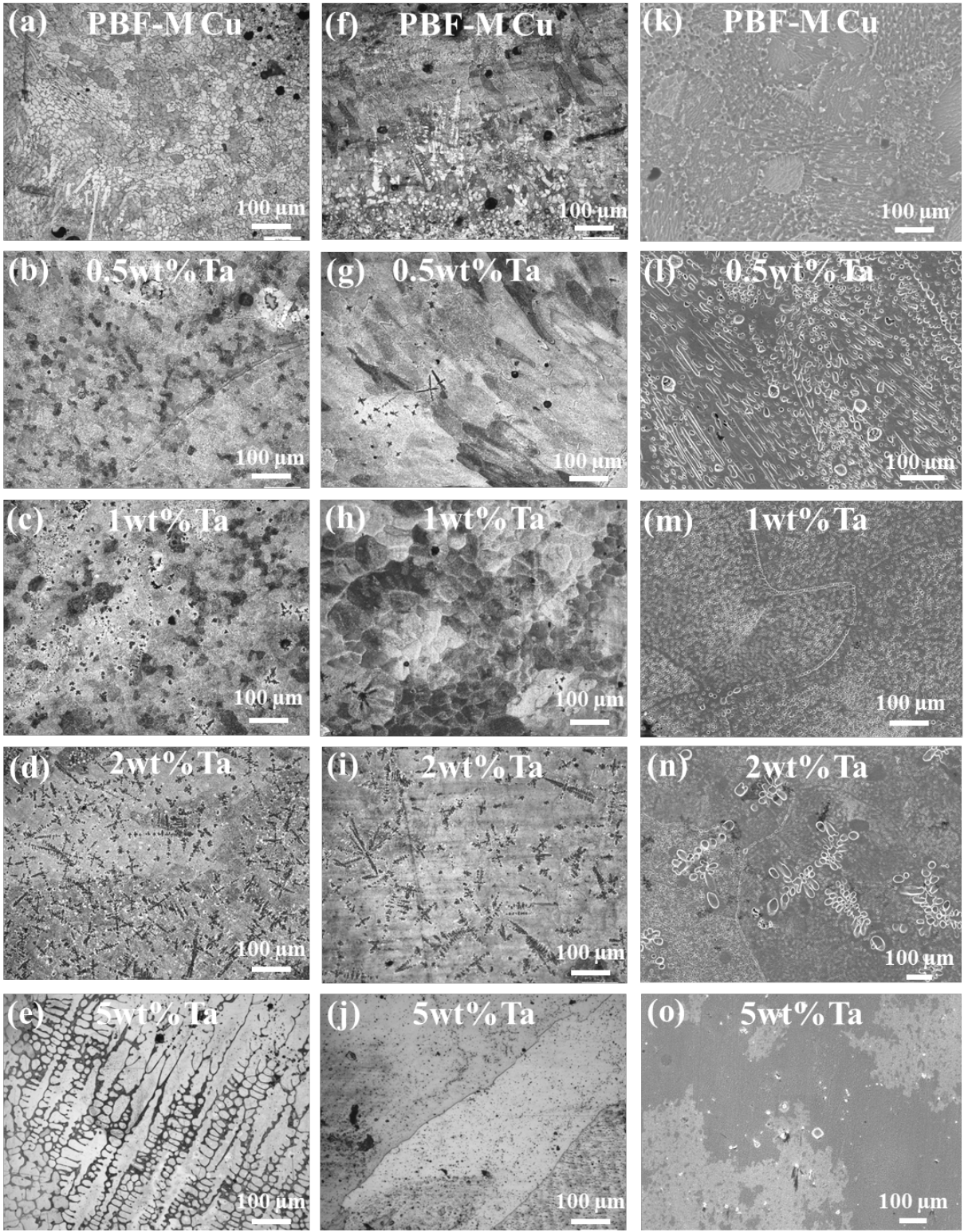

**Figure 1**: **(a)-(e)** Longitudinal micrograph of with PBF-M Cu, 0.5wt%Ta, 1wt%Ta, 2wt%Ta and 5wt%Ta, respectively, **(f)-(j)** Transverse micrograph of with PBF-M Cu, 0.5wt%Ta, 1wt%Ta, 2wt%Ta and 5wt%Ta, respectively, **(k)-(o)** SEM micrograph of PBF-M Cu, 0.5wt%Ta, 1wt%Ta, 2wt%Ta and 5wt%Ta, respectively.

The dispersion of Ta within the Cu matrix was examined using an HR-TEM. **Figure 2(a)** illustrates the presence of Ta particles with an approximate size of 600 nm, clearly delineating a phase boundary with the Cu matrix. In **Figure 2(b)**, a dark field TEM micrograph reveals the presence of Ta and its BCC phase, further confirmed through the selected area spot pattern shown in **Figure 2(c)**. Furthermore, a high-resolution transmission electron microscopy (HR-TEM) analysis of a Ta particle was performed, confirming the presence of the Ta (110) crystal plane, as shown in **Figure 2(d-e)**. The larger Ta particles exhibit smaller grains surrounding them, resulting in grayscale variation due to their misalignment with the zone axis. Consequently, the sample was analyzed using a Scanning Transmission Electron Microscope (STEM) in High Angle Annular Dark Field (HAADF) mode. The distinct contrast between Cu and Ta, facilitated by their significant atomic weight difference, can be observed in the HAADF image displayed in **Figure 2(f)**. One interesting phenomenon this study shows is that the initial Ta powder size used in the pellet decreased from the micrometer to nanometer range after the PBF-M process. During the sudden interaction of the laser with the powder mixture, the Cu matrix melts, but Ta particles experience fragmentation, as Ta possesses a higher temperature and remains unmelted during the PBF-M process. The fragmentation of Ta within the liquid Cu occurs during the melting and solidification process due to stress induced by high-intensity laser[40]. This fragmentation results in smoother boundaries of Ta nanoparticles, which directly cause strain on the Cu matrix. This phenomenon contributes to the increased mechanical strength of the Cu-Ta immiscible composite system. The Cu matrix has a high dislocation density due to the presence of Ta, which imparts the hardening in the materials[41], as shown in **Figure 2(g-h)**. The same boundary can be seen in the SEM micrographs presented in **Figure S3**.

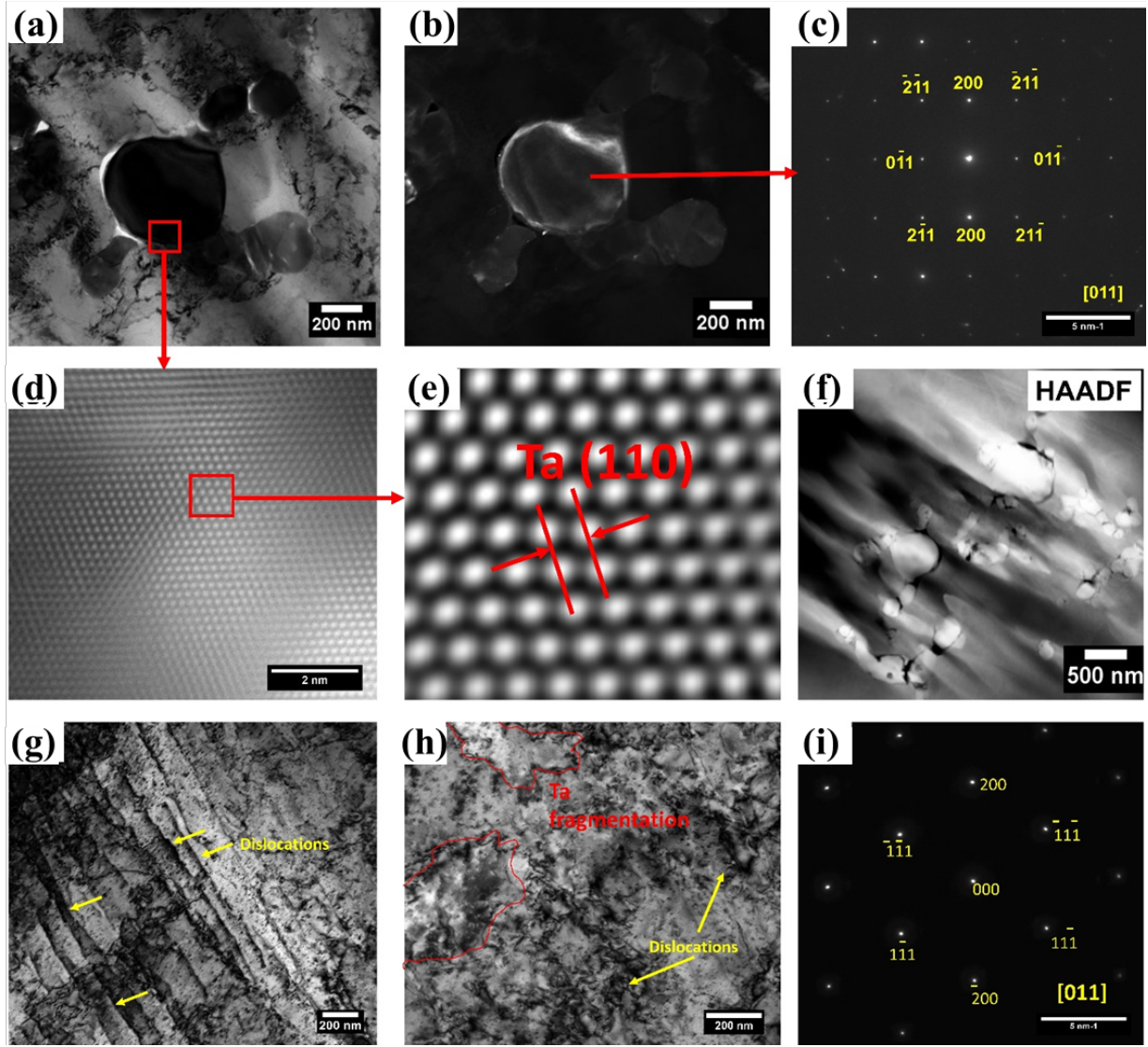

**Figure 2: (a)** brightfield TEM micrograph, **(b)** darkfield TEM micrograph showing the Ta particle, **(c)** spot diffraction pattern from the Ta particle, **(d)** HRTEM micrograph from Ta particle, **(e)** FFT filtered HRTEM image from image, **(f)** STEM, HAADF micrograph of Cu-Ta, **(g)** bright field TEM image dislocation rich Cu matrix, **(h)** bright field TEM micrograph with Ta and high-density dislocations in Cu matrix **(i)** selected area spot pattern from Cu matrix.

## 3.2. Composite Strength Analysis

XRD patterns of PBF-M Cu, 1wt%Ta, and 5wt%Ta are shown in **Figure 3(a)**. In addition, it can be seen that the left shift is for the main peak of Cu for Cu-Ta composites. This can be indicative of increased lattice parameters due to the PBF-M process-generated strain and Ta incorporation in the Cu lattice, which confirms the minor dissolution of Ta atoms in the matrix.

Such a result has been also found by Mula *et al.*[ref] and Kapoor *et al.* [ref] for Cu-Nb systems. They observed that Cu peaks move towards lower 2θ angles at higher Nb contents due to the expansion of the Cu lattice. This variation in Cu-Ta composites has occurred due to the roughening of Ta particles at the Cu and Ta interface, as shown in TEM micrographs, resulting in a lattice parameter increase. 5wt%Ta shows Ta peaks, whereas PBF-M Cu and 1wt%Ta do not show Ta peaks due to the absence and lower concentration of Ta, respectively. PBF-M Cu. 2wt%Ta shows oxide peaks in the sample, whereas 5wt%Ta does not show any oxide peak. The micro-Vickers hardness has been measured for the as-fabricated and heat-treated samples, and the results are presented in **Figure 3(b).** The increase in hardness observed in Cu-Ta composites can be attributed to two primary factors. Firstly, it is due to the dispersion of Ta particles within the matrix. Secondly, it results from the partial dissolution of Ta into the Cu matrix at the Cu-Ta interface or limited atomic level diffusion at the interface. The interface between Cu and Ta and the distribution of Ta in the Cu matrix play a crucial role in enhancing the microhardness. Therefore, initially, the hardness increases and this behavior indicates that nearly all the Ta retained its original powder shape within the Cu matrix. In addition, the formation of oxide can be taken into account as the other factor to increase the micro-hardness, as indicated by the XRD plot presented in **Figure 3(a)**. **Figure 3(c)** shows the stress-strain behavior of various Cu-Ta composites. These composites were tested at room temperature with a crosshead speed of 0.1 mm/min. The stress-strain plot of Cu-Ta composites with an FCC Cu matrix shows an increase in slope after reaching the ultimate tensile strength (UTS). Specifically, Cu-5 wt% Ta demonstrates a higher UTS value compared to other PBF-M-processed Cu-Ta composites studied in this research. Among Cu-Ta composites, the PBF-M Cu demonstrates a lower yield strength (YS), and Cu-5wt%Ta exhibits the highest YS, as shown in **Figure 3(d)**. The YS plot shows that Cu-2wt%Ta shows a 0.2% YS of 72.9 MPa, whereas Cu-5wt%Ta shows a higher YS of 80 MPa. The strain hardening coefficients (n) of the composites were evaluated and presented in **Figure S5**. As shown in **Figure S6**, the n value varies with Ta concentration in PBF-M of Cu, 0.5wt%Ta, 1wt%Ta, 2wt%Ta, and 5wt%Ta composites. The values of n for PBF-M Cu and 0.5wt%Ta are very close; beyond that, it decreases significantly from 0.5 to 0.2 for 1wt%Ta, 2wt%Ta. The plastic behavior of the composite (1wt%Ta and 2wt%Ta) is modified due to a change in dislocations interaction in the presence of Ta particles, which is further supported by MD simulations, which show surface roughening of Ta particles at the interface of Cu and Ta. Further addition of Ta, i.e., 5wt%Ta, results in the increase of n value, which can be due to the uniform distribution and nanometer size of Ta particles. This indicates that 5wt%Ta can withstand more deformation before necking and structural failure. Other PBF-M processed composites show smaller n values, indicating more prone to localized deformation and potential failure during plastic deformation, and it can be visualized in the stress-strain graph presented in **Figure 3(c)**. After the compression fracture morphology of composites was analyzed through SEM, the representative micrographs are presented in **Figures 4(a)** to **4(d).**

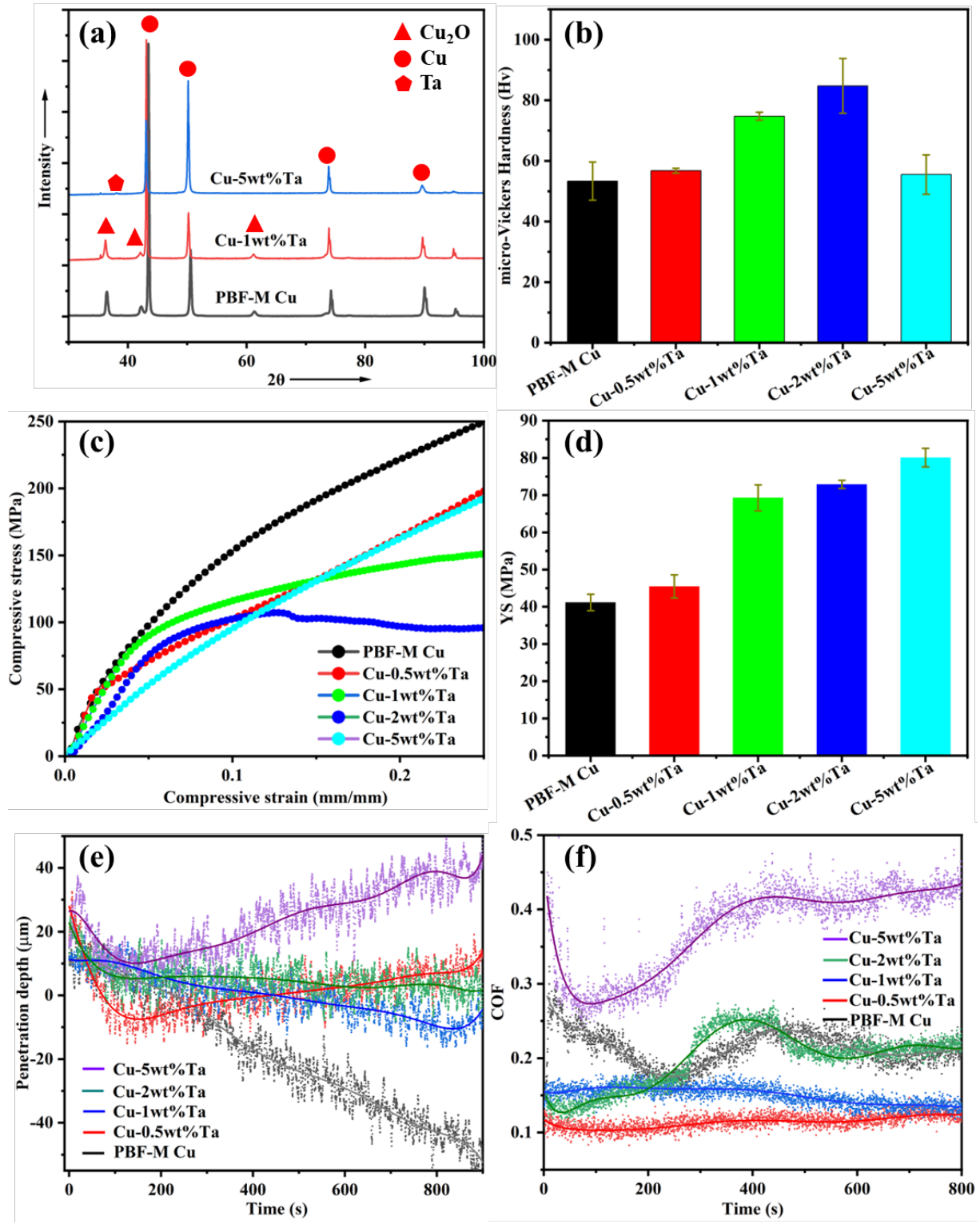

**Figure 3 (a)** X-ray diffraction pattern of Cu-Ta pellet Engineering stress-strain curve determined from Compression test performed at ambient temperature, **(b)** micro-Vickers hardness plot of PBF-M-processed Cu-Ta composites, **(c)** stress-strain diagram of Cu-Ta composites assesses from compression test, and **(d)** yield strength of different composites determined from stress-strain plot, **(e)** penetration depth plot of all investigated samples, **(f)** coefficient of friction plot.

To assess the tribological properties, Cu-Ta composites, along with PBF-M Cu, were subjected to fretting wear tests at room temperature using reciprocating motion. The results of the penetration depth, coefficient of friction (COF), and wear surface of pure Cu and Cu-Ta composites are shown in **Figures 4(c) and 4(d),** respectively. PBF-M Cu exhibits significant material degradation during the wear test compared to Cu-Ta composites. However, the formation of oxide on the surface helps the Cu-Ta composites remain unaffected by the abrasive nature of the counterpart. Cu-0.5wt%Ta shows a similar trend, but after a certain period of time, the removal of material from the composite stabilizes. On the other hand, Cu-5wt%Ta displays a higher penetration depth compared to the other investigated composites. This may be attributed to the presence of minor oxides in the composite or nano Ta structures working as abrasives. The COF of 0.5wt%Ta exhibits the minimum value, which correlates to the observed penetration depth for this composite. In contrast, Cu-5wt%Ta shows the maximum COF, leading to the highest penetration depth among the composites. This behavior may be attributed to the higher Ta content present in the composite compared to the other composites. Moreover, all the composites, including pure Cu, exhibit delamination during the tribological test, as shown in **Figure 4(e)-(f)**. Thermal conductivity data was obtained through an InfraRed camera setup, and the changes in temperature over time are presented in Supplementary Information **Figure S6**. 5wt%Ta shows a temperature difference($\Delta T$) of 8.3 °C at a time frame of 40 sec, whereas PBF-M Cu shows a $\Delta T$ of 5.9 °C at the same time frame. Thus, the difference in $\Delta T$ in both the samples suggests that PBF-M Cu is exhibiting a higher thermal conductivity value in comparison to 5wt%Ta. Thermal conductivity values are affected by the second phase and its distribution, grain boundary/interfaces, defects, etc. Although Cu shows higher thermal conductivity than Ta at 300K[42], the heat transfer across the particle-matrix interface is hindered by a smooth layer at the interface of Cu-matrix and Ta particles, as suggested and shown in TEM analysis (**Figure 2**). The importance of the particle-matrix interface in heat transfer is clear from CFD simulation. Although the interface between Cu and Ta particles is still free of any defects, the lower thermal conductivity of Ta particles will delay heat flow inside the matrix.

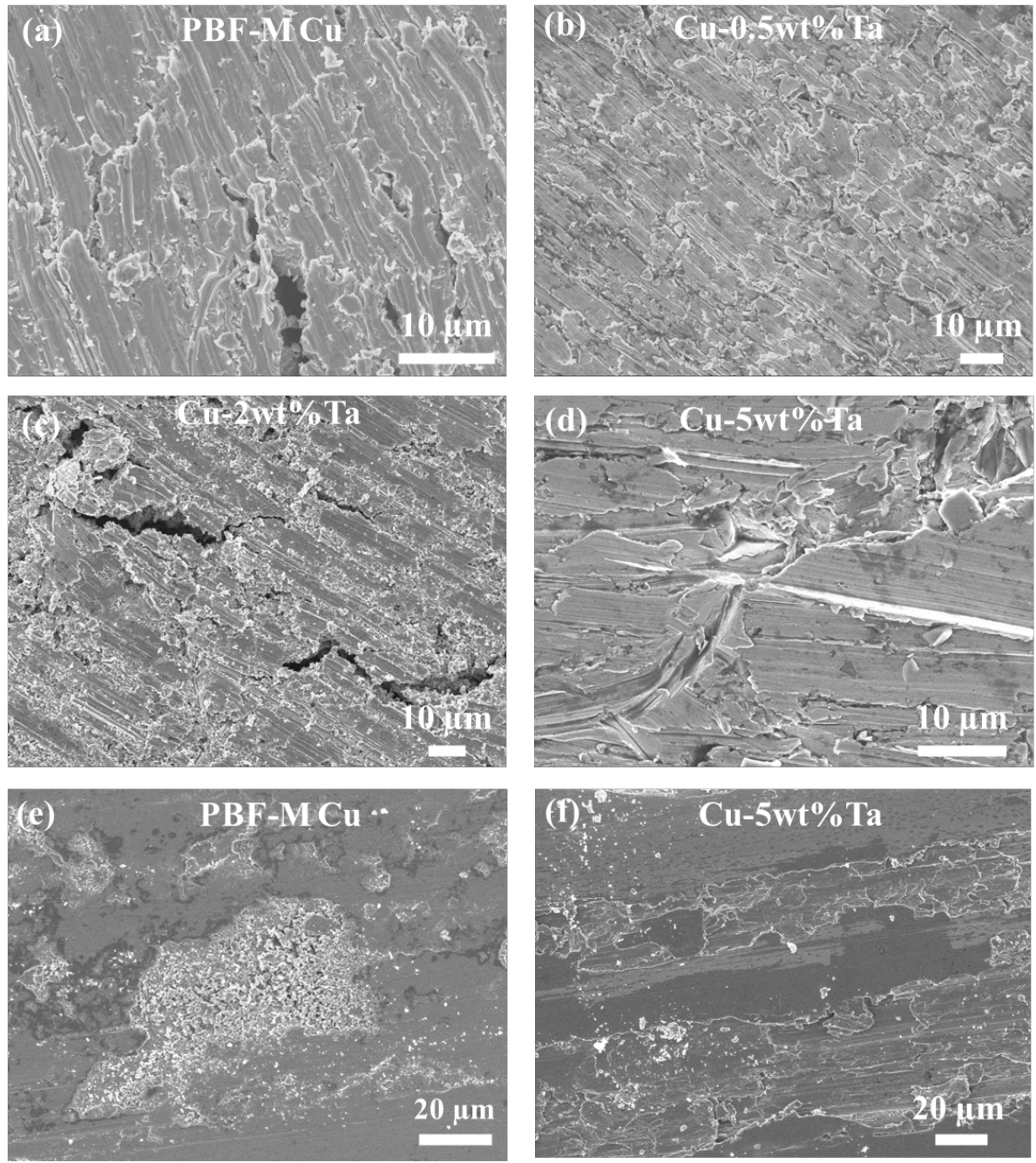

**Figure 4 (a)-(d)** SEM fracture morphology of compressed Cu-Ta composite in SE mode, **(e)-(f)** SEM wear surface morphology of PBF-M Cu and Cu-5wt%Ta respectively for comparison.

## 3.3. Modeling results

Energy dissipation is the main subject in laser-based melting; therefore in order to understand the melting behavior and microstructural evolution, we have performed coupled CFD-DEM simulation to gain insight into the melting process under a focus laser beam and quantify the heat flow/distribution and fluid flow in the melt pool during PBF-M processing. It is difficult to quantify these parameters using an experimental setup due to the rapid (µm-ms laser beam-material interaction times) nature of the PBF-M process. It provides temporal and spatial distribution of temperature along with associated rapid thermo-kinetics. The simulation of laser melting was carried out for the two configurations of the Cu-Ta pellet (2wt%Ta and 5wt%Ta) using the same laser processing parameters.

**Figure 5(a)** shows the cross-section of the 5wt%Ta pellet configuration, displaying a random distribution of Ta particles throughout the pellet. The pellet was divided into four sections (L1, L2, L3, L4) across the laser transverse direction or depth, as illustrated in **Figure 5(a)**. The temperature changes over time and the rate of change of temperature (heating and cooling rates) over time were plotted for these four locations and presented in **Figures 6(a)** and **6(b)**, respectively. The maximum temperature at L1 is significantly lower (below the melting point of Ta) than the maximum temperature at the laser center. This is due to the continuous heat loss from the top surface due to convection and radiation processes. Further, as we move downwards, convection and radiation from the curved surfaces of the cylinder take away most of the heat and conduction from the top subsides. At location L1, frequent oscillations are observed during the heating and cooling phases. These oscillations may be attributed to the significant difference between the thermal conductivities of Cu and Ta. The estimated heating and cooling rates from the simulation were $10^7$ K/s, indicating that Ta particles would have a short time window for redistribution within the melt pool and would get solidified in the Cu-matrix. In addition, the predicted peak temperature reached above the melting temperature of Cu while remaining well below the melting temperature of Ta. With a maximum heating rate of 9.72 x 10$^6$ K/s, the conduction velocity decreases as the heat gets conducted from the top surface and a Ta particle interacts.

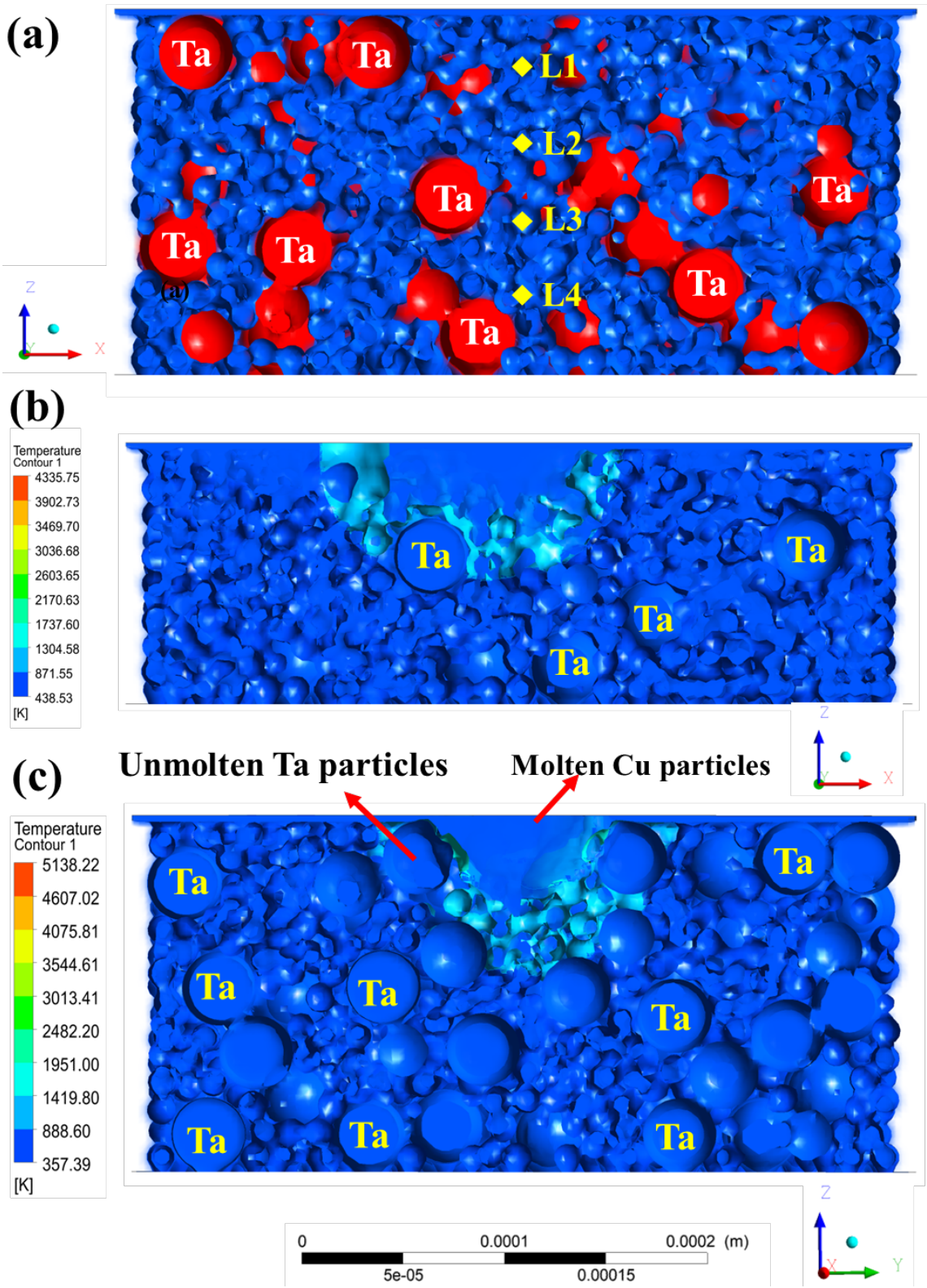

**Figure 5: (a)** The cross-section of the Cu-Ta pellet with 5wt%Ta showing larger Ta particles (in red) surrounded by smaller Cu particles (in blue). The four locations (uniformly spaced)

across the depth at which the temperature variations and the rate of change of temperature with time are tracked **(b)** Temperature contours at the longitudinal cross-section for 2wt%Ta, **(c)** The temperature contours at the cross-section in a longitudinal direction with a solidified zone of Cu around the intact Ta particles for 5wt%Ta.

The temperatures at locations L2, L3, and L4 are lower than those at location L1 due to the continuous solidification of the metals near the top and the superheat not being enough to reach the bottom. This can also be seen in **Figures 6(c) and 6(d)**, which show the temperature contours of the cross-section for the direction parallel to the laser motion. The melting pool's temperature distribution has been plotted on iso-volume fractions of copper. Since the modeling considers three phases (Cu, Ta, and air), the temperature contours were plotted on one of the metallic phases. This makes the contours discontinuous in appearance. The depth up to which the effect of laser heat can be seen is $\delta_1 = 0.0908$ mm (90.8 μm) for 5wt%Ta. Up to this distance, Cu particles are molten as the temperature is well above the melting point of Cu, but Ta particles are not molten as their melting point is higher than this temperature value. Similarly, in the case of the Cu-2wt%Ta, the laser heat affected the pellet in-depth $\delta_2 = 0.0387$ mm (38.7 μm), as shown in **Figure 6(c)**. This difference might be attributed to the differences in particle distribution in both cases. The maximum temperatures at the top surface of the pellets are 4335.75 K (2wt%Ta) and 5138.22 K (5wt%Ta). **Figure S9(a)** shows the solidified morphology, where copper particles are completely molten and surrounded by unmolten Ta particles. The temperature of the Cu melt pool ranges from 1304 K to 2170 K for the 2%Ta pellet and from 1419 K to 2482 K for the 5%Ta case. The lower thermal conductivity of Ta can retain the heat for maximum time, and with increment of Ta, the melt pool for 5wt%Ta remains at a higher temperature for maximum time. This is the reason behind the larger grain size observed for 5wt%Ta, as shown in **Figure 1**. While we have not considered the effect of recoil in the current modeling study, the maximum temperatures inside the melt pool are not enough to cause evaporation of the molten copper as these temperatures (2170 K and 2482 K) are below the evaporation point (2868 K). Therefore, the keyhole caused by the recoil will not be observed.

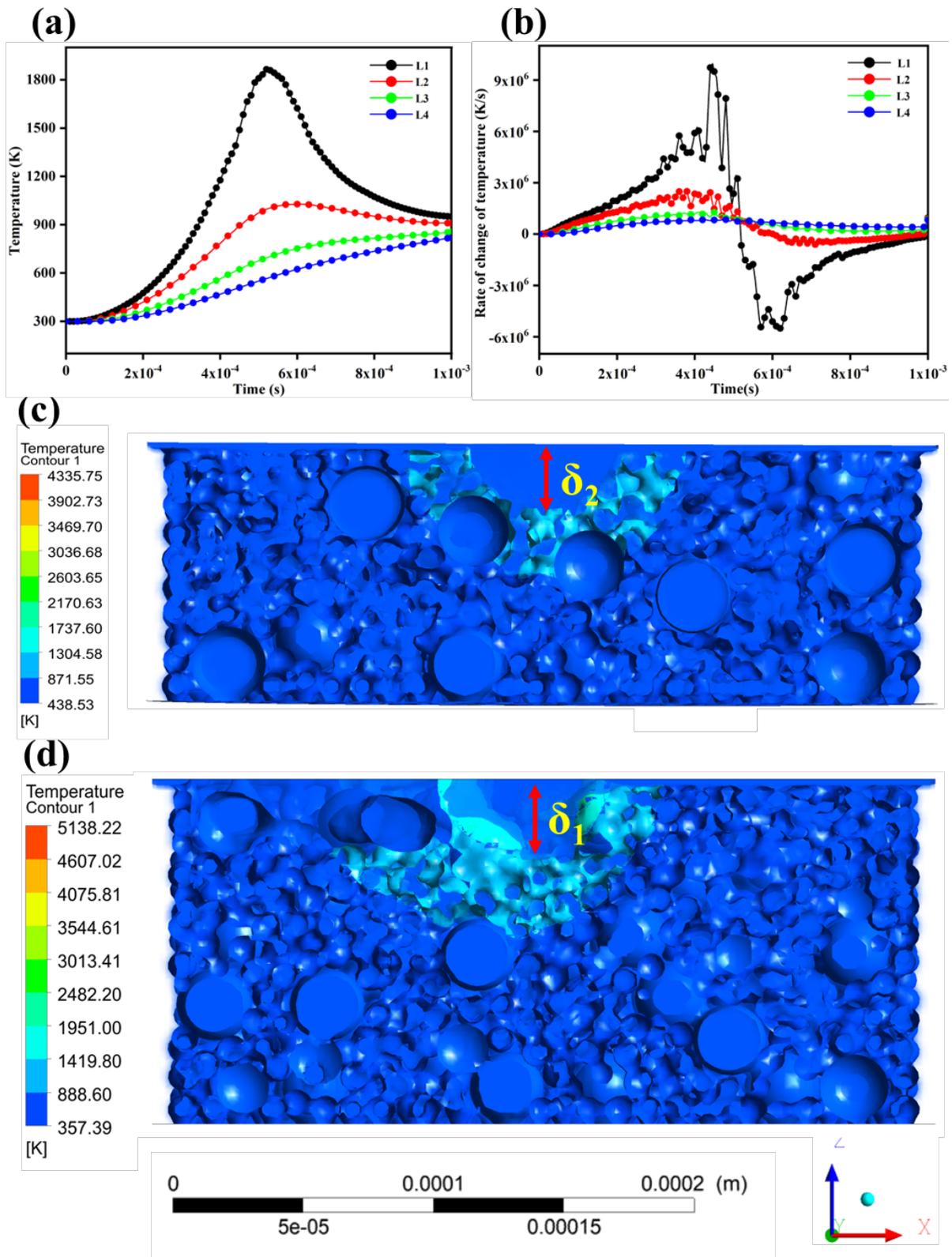

**Figure 6**: **(a)** The variation of temperature at the four locations with time, **(b)** The variation of heating and cooling rates at the four locations, **(c)** Temperature contour at the cross-section of the 2wt%Ta pellet in a direction parallel to the laser motion, and **(d)** The temperature contours at the cross-section of the pellet in longitudinal direction with maximum depth up to which melting of Cu particles has occurred for 5wt%Ta.

The real size of the pellets used in the experiments is in the order of millimeters. However, modeling such a large pellet would have increased the number of cells to the extent that it would be computationally costly. To accommodate this difference, we normalized the energy density. The melt pool depth observed in our study is lower due to the normalization used to scale down the size of the pellet. Also, the overall zone of influence of the laser, in this case, is smaller as there are more Ta particles near the top surface. The volume just below the laser has properly molten, and the completely molten Cu particles have formed a fusion zone around the Ta particles for 2wt%Ta and 5wt%Ta as inferred from Supporting Information **Figures S9(b)** and **S9(c),** respectively. These unmolten Ta particles in the solidified matrix of Cu increase the specific strength of the sample as shown in our experimental results. **Figures S8(a)** and **S8(b)** show the evolution of the surface of the re-solidified track with time as the laser progresses for 2wt%Ta and 5wt%Ta, respectively. As the injected laser energy starts melting the top surface and heat propagates downwards, the temperature rises quickly, and Cu particles get entirely molten. In contrast, the Ta particles are unmolten, forming an isolated island feature. As the laser moves forward, there is growth in the melt pool width, and smaller particles near the edge of the melt pool are partially molten and form a fusion zone with the re-solidified track. The end of the simulation reveals a completely molten and then solidified track with some variations in the track height and small globules of unmolten Cu particles at the track edges. The solidified track width in the case of 2wt%Ta is w = 0.07606 mm (76.06 μm). Supporting Information **Figures S9(b)** and **S9(c)** show the contours of liquid fraction for Cu and Ta in the melt pool at a particular instant. As shown in **Figure S9(b)**, the Cu particles are completely molten and form a homogenous solidified volume underneath the laser, while some particles get fused (due to less heat). **Figure S9(c)** shows that the liquid fraction for the Ta particles is zero and they are entirely unmolten, as is expected due to the high melting point of Ta. The partially molten Cu particles fuse with Ta particles, forming a compacted solidified zone. **Figure S9(a)** shows the uniformly molten boundary with unmolten Ta particles embedded at the boundary of the pellet after the laser motion has been completed for a single track. This indicates that the pellets can be uniformly melted for multiple laser tracks due to remelting. This numerical modeling data is also consistent with the micro-Vickers hardness values, as seen in **Figure 2(b)**, and the stress-strain curve presented in **Figure 2(c)**.

However, these theoretical results did not provide insights into the fragmentation of Ta particles. Therefore, MD simulations were performed for a similar situation. The Ta particles, measuring 2.7 nm in size, were embedded within the Cu matrix and subsequently annealed at a high temperature of 1000 K. After annealing, the particle shape became asymmetric, increasing the roughness of the interface between the metals, but the size of the particle remained approximately the same. In the MD simulations, the heating and cooling cycle aimed to mimic the temporary increase in temperature caused by the laser. Upon annealing, the surface of the Ta particles transformed, becoming rough. In this interface region, some Ta atoms diffused into Cu, and some Cu atoms diffused into Ta, despite their immiscibility, as shown in **Figure 7(a-c)**. Still, even after extended annealing, the Ta atoms did not diffuse beyond the interface region into the Cu-matrix. Furthermore, this nanoscale model revealed no evidence of the Ta particle dividing into two or more sub-particles. We attribute this to the small particle size. Instead, the partial mixing at the interface resulted in a continuous increase in the fraction of Ta atoms in this region, and the system remained in this state upon cooling (**Figure 7(e)**). This phenomenon suggests that the Ta particles underwent an expansion to enhance their surface area, with small regions of the particle partially detached from its bulk, with the heating providing energy to offset the small but positive enthalpy of mixing ($\Delta H=2$ kJ/mol)[43]. However, due to the inherent properties of the Cu-matrix, atomic mixing between the two materials was hindered. In the experiments' larger scale and higher temperature, the detachment of complete particle regions is possible, as observations made in the TEM analysis determined that micron-sized Ta particles fragmented within the Cu matrix.

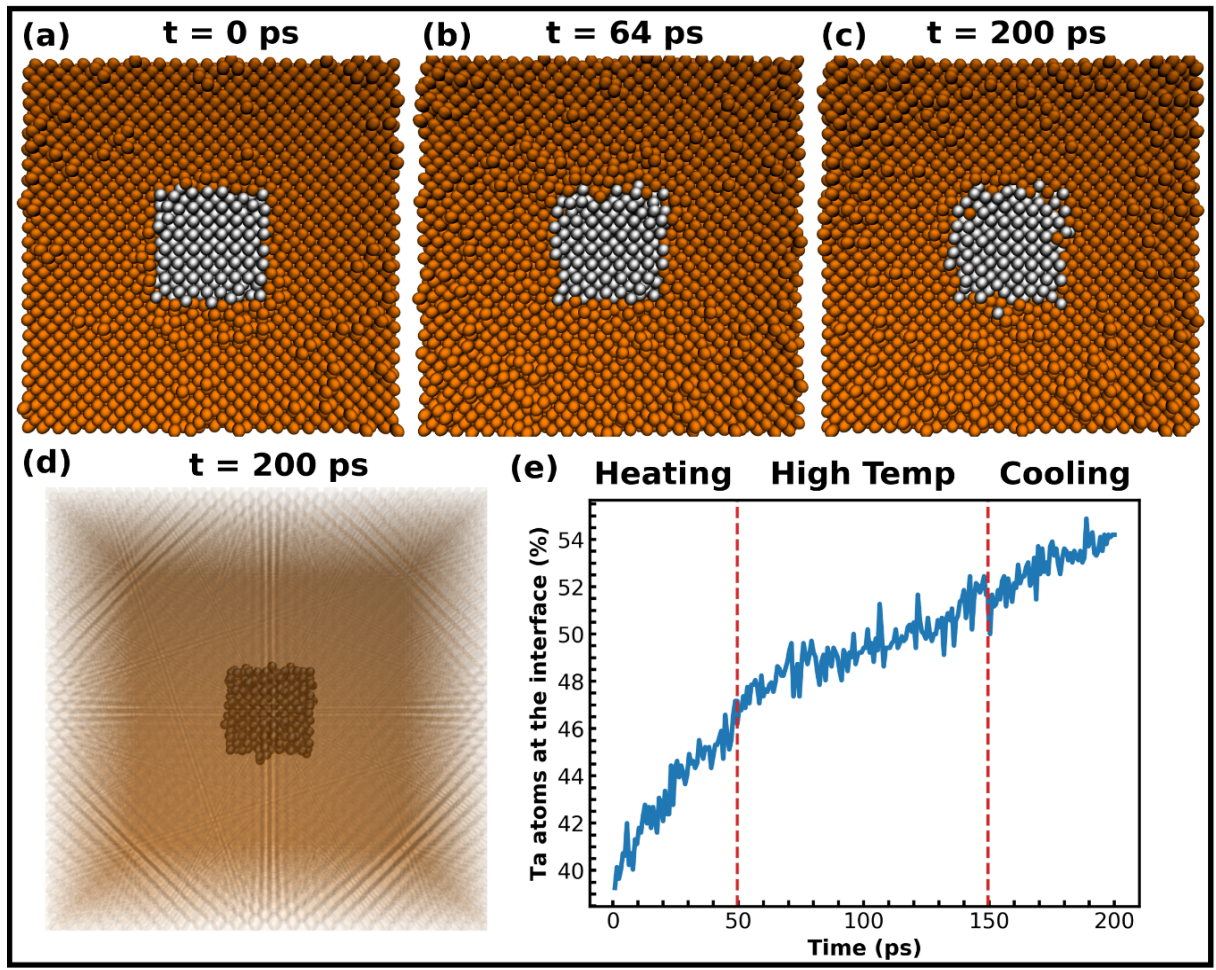

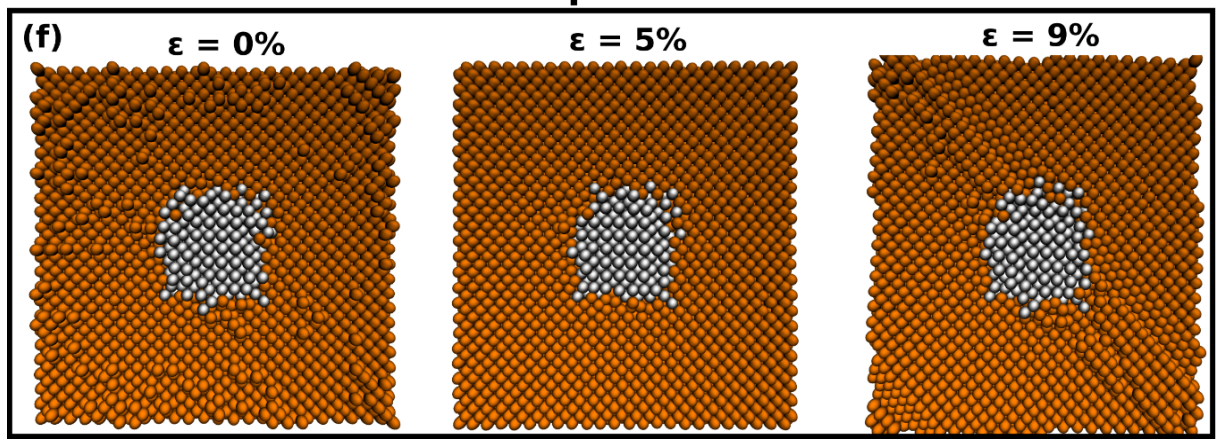

**Figure 7**: In the MD simulations, a Ta particle was embedded in a Cu matrix and annealed at 1000 K (727oC) for 200 ps. The representative MD snapshots show a cross-section of the system at (a) 0 ps, (b) 64 ps, and (c) 200 ps. The snapshot in (d) displays the entire system at the end of the 200 ps annealing simulation. The results in (e) show the time evolution of the fraction of Ta atoms at the Ta-Cu interface during the annealing process. We also simulated the compression of the system, and (f) displays the cross-section of compressed particles at different strain levels.

Similar observations were made in the TEM analysis, where micron-sized Ta particles within the Cu matrix were also found to have fragmented. After the annealing process was completed, the resulting structure was used as the starting point for a compression mechanical test. **Figure 7(f)** displays the structure before compression at 5% and 9% strain. Inspection of the initial configuration reveals that the heating-induced morphology changes remained in the Ta particle. In these simulations, we gradually applied strain to the Cu-Ta system to investigate how compression changed the region near a Ta particle. Results reveal that fracture in the simulations occurred at 9% strain. Examination of the final atomic configuration reveals that most deformation occurred in the Copper matrix. We also observed further diffusion of the Ta atoms into the interface near the Ta particle. We also simulated a system with a spherical Ta particle embedded in a Cu matrix with a similar atomic composition. We discuss the results for this case in the Supporting Information (**Figure S10**). The results for the spherical and cubic particles are similar, except that the initial fraction of Ta atoms near the interface was larger for the spherical particle, and the variation in the fraction of Ta atoms at the interface was higher for the cubic particle. As in the experiments, we found that the strength improves when we increase the concentration of Ta in Cu, as shown in the Supplementary Information in **Figure S7**. Also, we can clearly see that the hardening behavior changes as we change the Ta concentration, as we have observed in our experiments. The change in the value of the exponent is due to the change in dislocation interaction in the presence of Ta particles, as observed in our TEM images. These observations suggest that the dislocations piling up at the interface of Cu and Ta plays a significant role in increasing the strength of the Cu-Ta composite with an increase in a Ta content.

## 4. Conclusions

Despite their thermodynamic immiscibility, the PBF-M technique has proven to be effective in creating Cu-Ta composites. In this process, Ta is uniformly dispersed throughout the Cu matrix, forming a system where these two elements coexist. Interestingly, the Ta component appears as fragmented nanoparticles, contributing to the material's strength by imposing strain on the Cu lattice. Fully atomistic molecular dynamics simulations supported the experimental data interpretation. This mixture of Cu-Ta composite is well-suited for additive manufacturing, as it allows the fabrication of objects with complex shapes through layer-by-layer deposition. Remarkably, after undergoing the melting, the composite shows fragmentation of Ta particle from an initial size of micron range to nanometer one. This phenomenon is largely attributed to the sudden laser interaction at the Cu-Ta interfaces. The PBF-M processed Cu-Ta composite exhibits significantly high yield strength, making it a desirable choice for various applications. The results from the CFD modeling have also supported observations from the experiments. The Ta particles are unmolten in a completely molten and solidified Cu matrix. The maximum heating rate observed from the simulations is 9.72 x 10⁶ K/s. The melt pool depths observed are $\delta_1$ = 0.0908 mm (90.8 µm). and $\delta_2$ = 0.0387 mm (38.7 µm) for the 5wt % and 2wt % Ta cases, respectively. It was also observed that at the given parameters, the surface morphology of the tracks is uniformly molten.

**Declaration of Competing Interest:** The authors declare that they have no known competing financial interests or personal relationships that could have appeared to influence the work reported in this paper.

**CrediT authorship contribution statement**

**Rakesh Das:** Data curation, Validation, Formal analysis, Methodology, Investigation, Writing – original draft. **Pawan Kumar Dubey:** Validation, Software, Data Curation, Writing - Original Draft, Investigation. **Raphael Benjamim de Oliveira**: Data Curation, Software, Validation, Writing - Original Draft. **Douglas S. Galvao**: Writing – review & editing. **Indranil Manna:** Project administration, Writing – review & editing. **Sameehan S. Joshi:** Writing – review & editing. **Peter Samora Owuor:** Writing - Review & Editing, **Leonardo Dantas Machado:** Software, Validation, Writing – review & editing. **Nirmal Kumar Katiyar**: Writing – review & editing. **Suman Chakraborty:** Supervision, Resources, Project administration, Conceptualization, Writing – review & editing. **Chandra Sekhar Tiwary**: Supervision, Resources, Project administration, Conceptualization, Validation, Writing – review & editing.

**Data availability:** Data will be available on request.

**Acknowledgment:** Authors would like to thank Department of Science and Technology of India. LDM and RBO acknowledge the Brazilian agencies CNPq and CAPES for the financial support. LDM would also like to thank the support of the High-Performance Computing Center at UFRN (NPAD/UFRN). D. S. G. acknowledges the Center for Computing Engineering and Sciences at Unicamp for financial support through the FAPESP/CEPID grant #2013/08293-7.

**References**

[1] C. Tan, Q. Shi, K. Li, K. Khanlari, X. Liu, Effect of oxygen content of tantalum powders on the characteristics of parts processed by laser powder bed fusion, International Journal of Refractory Metals and Hard Materials 110 (2023) 106008. https://doi.org/10.1016/j.ijrmhm.2022.106008.

[2] S.L. Sing, F.E. Wiria, W.Y. Yeong, Selective laser melting of titanium alloy with 50 wt% tantalum: Effect of laser process parameters on part quality, International Journal of Refractory Metals and Hard Materials 77 (2018) 120–127. https://doi.org/10.1016/j.ijrmhm.2018.08.006.

[3]  D.W. Yee, M.L. Lifson, B.W. Edwards, J.R. Greer, Additive Manufacturing of 3D-Architected Multifunctional Metal Oxides, Adv. Mater. 31 (2019) 1901345. https://doi.org/10.1002/adma.201901345.

[4]  T.M. Wischeropp, C. Emmelmann, M. Brandt, A. Pateras, Measurement of actual powder layer height and packing density in a single layer in selective laser melting, Additive Manufacturing 28 (2019) 176–183. https://doi.org/10.1016/j.addma.2019.04.019.

[5]  S. Ma, Z. Shang, A. Shang, P. Zhang, C. Tang, Y. Huang, C.L.A. Leung, P.D. Lee, X. Zhang, X. Wang, Additive manufacturing enabled synergetic strengthening of bimodal reinforcing particles for aluminum matrix composites, Additive Manufacturing 70 (2023) 103543. https://doi.org/10.1016/j.addma.2023.103543.

[6]  L. Constantin, N. Kraiem, Z. Wu, B. Cui, J.-L. Battaglia, C. Garnier, J.-F. Silvain, Y.F. Lu, Manufacturing of complex diamond-based composite structures via laser powder-bed fusion, Additive Manufacturing 40 (2021) 101927. https://doi.org/10.1016/j.addma.2021.101927.

[7]  L. Yan, Y. Chen, F. Liou, Additive manufacturing of functionally graded metallic materials using laser metal deposition, Additive Manufacturing 31 (2020) 100901. https://doi.org/10.1016/j.addma.2019.100901.

[8]  J. Zhang, C. Chen, G. Wang, Z. Geng, D. Li, Y. Wu, K. Zhou, Mechanical behaviors of tantalum scaffolds with node optimization fabricated by laser powder bed fusion, International Journal of Refractory Metals and Hard Materials 124 (2024) 106837. https://doi.org/10.1016/j.ijrmhm.2024.106837.

[9]  M. Wang, R.S. Averback, P. Bellon, S. Dillon, Chemical mixing and self-organization of Nb precipitates in Cu during severe plastic deformation, Acta Materialia 62 (2014) 276–285. https://doi.org/10.1016/j.actamat.2013.10.009.

[10]  D. Zhang, D. Qiu, M.A. Gibson, Y. Zheng, H.L. Fraser, D.H. StJohn, M.A. Easton, Additive manufacturing of ultrafine-grained high-strength titanium alloys, Nature 576 (2019) 91–95. https://doi.org/10.1038/s41586-019-1783-1.

[11]  L. Constantin, Z. Wu, N. Li, L. Fan, J.-F. Silvain, Y.F. Lu, Laser 3D printing of complex copper structures, Additive Manufacturing 35 (2020) 101268. https://doi.org/10.1016/j.addma.2020.101268.

[12]  S.D. Jadhav, P.P. Dhekne, S. Dadbakhsh, J.-P. Kruth, J. Van Humbeeck, K. Vanmeensel, Surface Modified Copper Alloy Powder for Reliable Laser-based Additive Manufacturing, Additive Manufacturing 35 (2020) 101418. https://doi.org/10.1016/j.addma.2020.101418.

[13]  C. Wei, L. Liu, Y. Gu, Y. Huang, Q. Chen, Z. Li, L. Li, Multi-material additive-manufacturing of tungsten - copper alloy bimetallic structure with a stainless-steel interlayer and associated bonding mechanisms, Additive Manufacturing 50 (2022) 102574. https://doi.org/10.1016/j.addma.2021.102574.

[14]  T. Venugopal, K. Prasad Rao, B.S. Murty, Mechanical and electrical properties of Cu–Ta nanocomposites prepared by high-energy ball milling, Acta Materialia 55 (2007) 4439–4445. https://doi.org/10.1016/j.actamat.2007.04.025.

[15]  K.A. Darling, E.L. Huskins, B.E. Schuster, Q. Wei, L.J. Kecskes, Mechanical properties of a high strength Cu–Ta composite at elevated temperature, Materials Science and Engineering: A 638 (2015) 322–328. https://doi.org/10.1016/j.msea.2015.04.069.

[16]  L.-M. Luo, X.-P. Ding, W.-Z. Xu, C.-Y. Wang, Y.-Q. Qin, Y.-C. Wu, Strengthening mechanisms and research progress in the W/Cu interfaces, International Journal of Refractory Metals and Hard Materials 125 (2024) 106900. https://doi.org/10.1016/j.ijrmhm.2024.106900.

[17] F.T.N. Vüllers, R. Spolenak, From solid solutions to fully phase separated interpenetrating networks in sputter deposited "immiscible" W–Cu thin films, Acta Materialia 99 (2015) 213–227. https://doi.org/10.1016/j.actamat.2015.07.050.

[18] K.A. Darling, E.L. Huskins, B.E. Schuster, Q. Wei, L.J. Kecskes, Mechanical properties of a high strength Cu–Ta composite at elevated temperature, Materials Science and Engineering: A 638 (2015) 322–328. https://doi.org/10.1016/j.msea.2015.04.069.

[19] S. Farag, I. Konyashin, B. Ries, The influence of grain growth inhibitors on the microstructure and properties of submicron, ultrafine and nano-structured hardmetals – A review, International Journal of Refractory Metals and Hard Materials 77 (2018) 12–30. https://doi.org/10.1016/j.ijrmhm.2018.07.003.

[20] T. Venugopal, K. Prasad Rao, B.S. Murty, Mechanical and electrical properties of Cu–Ta nanocomposites prepared by high-energy ball milling, Acta Materialia 55 (2007) 4439–4445. https://doi.org/10.1016/j.actamat.2007.04.025.

[21] T. Frolov, K.A. Darling, L.J. Kecskes, Y. Mishin, Stabilization and strengthening of nanocrystalline copper by alloying with tantalum, Acta Materialia 60 (2012) 2158–2168. https://doi.org/10.1016/j.actamat.2012.01.011.

[22] R. Das, N.K. Katiyar, S. Sarkar, S. Sarkar, V. Mishra, C.S. Tiwary, Engineering the Interface of Cu-hBN Immiscible System Using 3D Printing To Enhance Mechanical and Thermal Properties, ACS Appl. Eng. Mater. 2 (2024) 1234–1244. https://doi.org/10.1021/acsaenm.3c00722.

[23] A.P. Mouritz, ed., 4 - Strengthening of metal alloys, in: Introduction to Aerospace Materials, Woodhead Publishing, 2012: pp. 57–90. https://doi.org/10.1533/9780857095152.57.

[24] K.M. Mannan, K.R. Karim, Grain boundary contribution to the electrical conductivity of polycrystalline Cu films, J. Phys. F: Met. Phys. 5 (1975) 1687–1693. https://doi.org/10.1088/0305-4608/5/9/009.

[25] M.F. Francis, M.N. Neurock, X.W. Zhou, J.J. Quan, H.N.G. Wadley, E.B. Webb, Atomic assembly of Cu/Ta multilayers: Surface roughness, grain structure, misfit dislocations, and amorphization, Journal of Applied Physics 104 (2008) 034310. https://doi.org/10.1063/1.2968240.

[26] I. Lazić, P. Klaver, B. Thijsse, Microstructure of a Cu film grown on bcc Ta (100) by large-scale molecular-dynamics simulations, Phys. Rev. B 81 (2010) 045410. https://doi.org/10.1103/PhysRevB.81.045410.

[27] T.P.C. Klaver, B.J. Thijsse, Molecular Dynamics simulations of Cu/Ta and Ta/Cu thin film growth, Journal of Computer-Aided Materials Design 10 (2003) 61–74. https://doi.org/10.1023/B:JCAD.0000036802.46424.ee.

[28] A.-S. Tran, Phase transformation and interface fracture of Cu/Ta multilayers: A molecular dynamics study, Engineering Fracture Mechanics 239 (2020) 107292. https://doi.org/10.1016/j.engfracmech.2020.107292.

[29] L. Lu, C. Huang, W. Pi, H. Xiang, F. Gao, T. Fu, X. Peng, Molecular dynamics simulation of effects of interface imperfections and modulation periods on Cu/Ta multilayers, Computational Materials Science 143 (2018) 63–70. https://doi.org/10.1016/j.commatsci.2017.10.034.

[30] H.R. Gong, B.X. Liu, Influence of interfacial texture on solid-state amorphization and associated asymmetric growth in immiscible Cu-Ta multilayers, Phys. Rev. B 70 (2004) 134202. https://doi.org/10.1103/PhysRevB.70.134202.

[31] G.P. Purja Pun, K.A. Darling, L.J. Kecskes, Y. Mishin, Angular-dependent interatomic potential for the Cu–Ta system and its application to structural stability of nano-crystalline alloys, Acta Materialia 100 (2015) 377–391. https://doi.org/10.1016/j.actamat.2015.08.052.

[32] J. Shi, J. Wang, X. Yi, Y. Lu, D. Hua, Q. Zhou, X. Fan, Nanoscratching-induced plastic deformation mechanism and tribology behavior of Cu/Ta bilayer and multilayer by a molecular dynamics study, Applied Surface Science 586 (2022) 152775. https://doi.org/10.1016/j.apsusc.2022.152775.

[33] M.S. Daw, M.I. Baskes, Embedded-atom method: Derivation and application to impurities, surfaces, and other defects in metals, Phys. Rev. B 29 (1984) 6443–6453. https://doi.org/10.1103/PhysRevB.29.6443.

[34] X.W. Zhou, R.A. Johnson, H.N.G. Wadley, Misfit-energy-increasing dislocations in vapor-deposited CoFe/NiFe multilayers, Phys. Rev. B 69 (2004) 144113. https://doi.org/10.1103/PhysRevB.69.144113.

[35] X.W. Zhou, H.N.G. Wadley, R.A. Johnson, D.J. Larson, N. Tabat, A. Cerezo, A.K. Petford-Long, G.D.W. Smith, P.H. Clifton, R.L. Martens, T.F. Kelly, Atomic scale structure of sputtered metal multilayers, Acta Materialia 49 (2001) 4005–4015. https://doi.org/10.1016/S1359-6454(01)00287-7.

[36] M.S. Daw, M.I. Baskes, Embedded-atom method: Derivation and application to impurities, surfaces, and other defects in metals, Phys. Rev. B 29 (1984) 6443–6453. https://doi.org/10.1103/PhysRevB.29.6443.

[37] X.W. Zhou, R.A. Johnson, H.N.G. Wadley, Misfit-energy-increasing dislocations in vapor-deposited CoFe/NiFe multilayers, Phys. Rev. B 69 (2004) 144113. https://doi.org/10.1103/PhysRevB.69.144113.

[38] A.P. Thompson, H.M. Aktulga, R. Berger, D.S. Bolintineanu, W.M. Brown, P.S. Crozier, P.J. In 'T Veld, A. Kohlmeyer, S.G. Moore, T.D. Nguyen, R. Shan, M.J. Stevens, J. Tranchida, C. Trott, S.J. Plimpton, LAMMPS - a flexible simulation tool for particle-based materials modeling at the atomic, meso, and continuum scales, Computer Physics Communications 271 (2022) 108171. https://doi.org/10.1016/j.cpc.2021.108171.

[39] G.P. Purja Pun, K.A. Darling, L.J. Kecskes, Y. Mishin, Angular-dependent interatomic potential for the Cu–Ta system and its application to structural stability of nano-crystalline alloys, Acta Materialia 100 (2015) 377–391. https://doi.org/10.1016/j.actamat.2015.08.052.

[40] M. Scisciò, M. Barberio, C. Liberatore, S. Veltri, A. Laramée, L. Palumbo, F. Legaré, P. Antici, Analysis of induced stress on materials exposed to laser-plasma radiation during high-intensity laser experiments, Applied Surface Science 421 (2017) 200–204. https://doi.org/10.1016/j.apsusc.2016.12.004.

[41] S. Zhou, M. Xie, C. Wu, Y. Yi, D. Chen, L.-C. Zhang, Selective laser melting of bulk immiscible alloy with enhanced strength: Heterogeneous microstructure and deformation mechanisms, Journal of Materials Science & Technology 104 (2022) 81–87. https://doi.org/10.1016/j.jmst.2021.06.062.

[42] T.M. Tritt, ed., Thermal Conductivity, Springer US, 2004. https://doi.org/10.1007/b136496.

[43] K.A. Darling, A.J. Roberts, Y. Mishin, S.N. Mathaudhu, L.J. Kecskes, Grain size stabilization of nanocrystalline copper at high temperatures by alloying with tantalum, Journal of Alloys and Compounds 573 (2013) 142–150. https://doi.org/10.1016/j.jallcom.2013.03.177.